\documentclass[prl,twocolumn,showpacs,preprintnumbers,amsmath,amssymb]{revtex4}
\usepackage{amsfonts}
\usepackage{bbm}
\usepackage{mathrsfs}
\usepackage{amsmath}
\usepackage[dvips]{graphics,color}
\usepackage{graphicx}
\usepackage{epstopdf}
\usepackage{dcolumn}
\usepackage{bm}
\usepackage{longtable}
\usepackage{graphics}
\usepackage{amssymb}
\usepackage{xspace}
\usepackage{epsfig}
\usepackage{subfigure}
\usepackage{dcolumn}
\makeatletter

\newcommand{\Rmnum}[1]{\expandafter\@slowromancap\romannumeral #1@}
\makeatother

\begin{document}
\title{Distinguishing Spontaneous Quantum Hall States in Graphene Bilayers}
\author{Fan Zhang}\email{zhangfan@physics.utexas.edu}
\author{A.H. MacDonald}
\affiliation{Department of Physics, University of Texas at Austin, Austin TX 78712, USA}
\date{\today}
\begin{abstract}
Chirally stacked $N$-layer graphene with $N\ge2$ is susceptible to a variety of distinct broken symmetry states in which each spin-valley flavor spontaneously transfers charge between layers.  In mean-field theory the neutral bilayer ground state is a layer antiferromagnet (LAF) state that has opposite spin-polarizations in opposite layers.  In this Letter we analyze how the LAF and other competing states are influenced by Zeeman fields that couple to spin and by interlayer electric fields that couple to layer pseudospin, and comment on the possibility of using response and edge state
signatures to identify the character of the bilayer ground state experimentally.
\end{abstract}
\maketitle

{\em Introduction.}---Bilayer graphene\cite{graphene_review,AB_McCann} and its thicker $N$-layer cousins, chirally (or ABC)
stacked multilayers\cite{ft1,ABC_Zhang,ABC_McCann,Chiral_Min,SQH_Zhang}, have attracted considerable
theoretical\cite{SQH_Zhang,MF_Min,RG_Zhang,MF_Levitov,SQH_Levitov,MF_Jung,RG_Vafek,RG_Falko} and experimental\cite{Yacoby1,Yacoby2,AB_Schonenberger,AB_Lau,AB_Lau1,ABC_Lau} attention because of their susceptibility to broken symmetries that are accompanied by large momentum space Berry curvatures and different types of topological order.  In a continuum model mean-field theory, the ground state is\cite{MF_Min,ft2} an Ising layer-pseudospin ferromagnet in which each spin-valley flavor is\cite{MF_Min,RG_Zhang,MF_Levitov} layer polarized.  The quasiparticle Hamiltonian in these states develops mass gaps that change the character of the wavefunctions at small momentum and produce\cite{SQH_Zhang,SQH_Levitov} Berry curvature.  The integral of Berry curvature over a suitably defined region of momentum space near a given valley is nearly exactly quantized at $\pm 2\pi$.  This property can be interpreted\cite{ft3} as saying that each valley contributes $\pm e^2/h$ to the Hall conductivity with a sign that reverses with valley index and with the sense of layer polarization.  States with total Hall conductivity $ I e^2/h$ evolve\cite{dagim}
smoothly into quantum Hall ferromagnets with $\nu=I$ in the presence of a perpendicular magnetic field.
\newline\indent
When spin is ignored only two different types of states can be distinguished, ones in which the $K$ and $K'$ valleys are layer polarized in the opposite sense producing a quantum anomalous Hall (QAH) state\cite{SQH_Zhang,SQH_Levitov,AH_Haldane} with broken time reversal ($\mathcal{T}$) symmetry and orbital magnetization\cite{SQH_Zhang}, and ones in which the two valleys have the same sense of layer polarization
producing an inversion ($\mathcal{I}$) symmetry breaking quantum valley Hall (QVH) state\cite{SQH_Zhang,SQH_Levitov} with zero total Hall conductivity.  When spin is included, there are three distinct states with no overall layer polarization as summarized in Table~\ref{table:one}: i) a QAH state with Hall effect contributions of the same sign for opposite spins, ii) a quantum spin Hall (QSH) state\cite{SQH_Zhang,SQH_Levitov,TI,QSH_Kane,QSH_Konig,QSH_Raghu} with opposite QAH signs for opposite spins, and iii) a LAF state\cite{SQH_Zhang} that has QVH states with opposite layer polarization signs for opposite spins.  Among these possibilities, lattice mean-field theory calculations\cite{MF_Jung} suggest that inter-valley exchange weakly favors QVH states in the spinless case and LAF states in the spinful case.
In this Letter we analyze how all three states respond to Zeeman coupling to their spin and to electric-field coupling to their layer pseudospin degrees-of-freedom.  We find that the Zeeman field response distinguishes QAH states from QSH and LAF states.  In the LAF, the Zeeman field induces a non-collinear spin state in which the components of the spin-density perpendicular to the field are opposite in opposite layers, while those along the field direction grow smoothly with field strength and are identical.  The three states respond similarly to an electric field between the layers,
which can induce first order transitions at which the total layer polarization jumps.
\addtolength{\textfloatsep}{-0.2in} 
\begin{table}[b]
\caption{Summary\cite{SQH_Zhang} of spin-valley layer polarizations ($t$ or $b$), broken symmetries, charge (C) and spin (S) Hall conductivities ($e^2/h$ units) and insulator types for the three distinct states with no overall layer polarization.}
\newcommand\T{\rule{0pt}{3.1ex}}
\newcommand\B{\rule[-1.7ex]{0pt}{0pt}}
\begin{scriptsize}
\centering
\begin{tabular}{c c c c | c | c | c | c }
\hline\hline $K\uparrow$\; & $K\downarrow$\; & $K'\uparrow$\; & $K'\downarrow$\;
& \;$\sigma^{\rm (S)}$\; & $\sigma^{\rm (C)}$ ($I$)& Broken Symm. & Insulator\T\\[3pt]
\hline
 t & t & b & b  &  $0$  & $2N$ & $\mathcal{T}$,\;\;$\mathcal{Z}_{2}$\quad\quad & QAH \T \\[3pt]
 t & b & t & b  &  $0$  & $0$  & \;\,$\mathcal{T}$,\;\;$\mathcal{SU}(2)$ & LAF \T \\[3pt]
 t & b & b & t  &  $2N$ & $0$  & $\mathcal{Z}_2$,\;\;$\mathcal{SU}(2)$ & QSH \T \\[3pt]
\hline\hline
\end{tabular}
\end{scriptsize}
\label{table:one}
\end{table}

There is already some suggestive experimental evidence for spontaneous quantum Hall states in graphene
multilayers that is consistent with mass gaps $\Delta \sim 2-8$ meV in recent studies of suspended bilayers\cite{Yacoby1,Yacoby2,AB_Schonenberger,AB_Lau,AB_Lau1} and trilayers\cite{ABC_Lau} .  Since the gaps are seen\cite{Yacoby2,AB_Lau,ABC_Lau} only at temperatures well below $\Delta/k_B$ they appear to be of many-body origin.
Moreover, measurements of bilayers in a perpendicular magnetic field $B$ appear to show that both $\nu=\pm 4$ and $\nu=0$ quantum Hall states can persist to zero-magnetic field\cite{AB_Schonenberger,AB_Lau}, implying that spontaneous quantum Hall states with total Hall conductivity quantum number $I=0,4$ can be stabilized by interactions at $B=0$.

{\em Continuum model mean-field theory.}---In single-layer graphene the band dispersion remains linear over a broad range of energy surrounding the charge neutrality point.  When graphene's honeycomb layers are chirally stacked only two sublattice sites, one located in the top layer and one in the bottom layer, are not connected to near-neighbors in other layers and are therefore relevant at low energies.  Hopping between these sites, {\em e.g.} from top ($t$) $A$ to bottom ($b$) $B$, becomes an $N$-step process, leading to two remarkably flat bands with $\pm k^{\rm N}$ dispersion and
layer pseudospin chirality $N$\cite{ABC_Zhang,SQH_Zhang}.  These unique band features are encoded in the low-energy $\!\bf{k}\!\cdot\!\bf{p}\!$
Hamiltonian given below.  Because of the flat bands and the large pseudospin chirality, interactions become dominant at low energies in few-layer\cite{ABC_Zhang,SQH_Zhang} chiral graphene.  In mean-field theory inversion symmetry is broken\cite{RG_Zhang} within each spin and
valley, leading in a contact interaction model to the following Hamiltonian:
\begin{subequations}
\begin{eqnarray}
\!\!\!{\mathcal H}^{\rm HF}&=&\sum_{{\bm k}\alpha\beta s s'}c^{\dag}_{{\bm k}\rm \alpha s}\big[h_0+h_{\rm H}+h_{\rm F}\big]
c_{{\bm k}\rm \beta s'}\,,\label{eq:hfH}\\
\!\!\!h_0 &=& \epsilon_{\bm k} \big[\cos(N \phi_{\bm k})\sigma_{\rm x}^{\rm \alpha\beta}+\sin(N \phi_{\bm k})\sigma_{\rm y}^{\rm \alpha\beta}\big]\delta_{\rm ss'}\,,\label{eq:bandH}\\
\!\!\!h_{\rm H} &=& \big[V_0 \Delta_{0} \delta^{\rm \alpha\beta} + V_{\rm z} \Delta_{\rm z} \sigma^{\rm \alpha\beta}_{\rm z}\big]\delta_{\rm ss'}\,,\\
\!\!\!h_{\rm F} &=& - \big[V_0 + V_{\rm z}\sigma^{\rm \alpha\alpha}_{\rm z}\sigma^{\rm \beta\beta}_{\rm z}\big]\Delta_{\rm \alpha s}^{\rm \beta s'}\,,
\end{eqnarray}
\end{subequations}
where $\epsilon_{\bm k}={(v_0 \hbar k)^N}/{(-\gamma_1)^{\rm N-1}}$ is the band dispersion, $V_{\rm 0,z}=(V_{\rm S} \pm V_{\rm D})/2$ denotes the average (difference) of intralayer and interlayer interactions, and $\Delta_{\rm \alpha s}^{\rm \beta s'}=A^{-1}\sum_{{\bm k}}\langle c^{\dag}_{{\bm k}\rm \beta s'}c_{{\bm k}\rm \alpha s}\rangle_{\rm f}$ must be determined self-consistently.  $\Delta_{\rm 0,z}$ is the density sum (difference) of the top and bottom layers.  $\cot\phi_{\bm k}=\tau_z k_{\rm x}/k_{\rm y}$ and $\tau_{\rm z}(\pm 1)$ labels valleys $K$ and $K'$.  The Pauli matrices ${\bm {\sigma}}$ act on the {\em which-layer} pseudospin and $s(\pm 1)$ denotes the real spin.  Because of the in-plane rotational symmetry of the continuum model, it is easy to verify that this mean-field Hamiltonian does not generate Hartree (H) or Fock (F) potentials that are off-diagonal in layer index.

We seek self-consistent solutions for the $N=2$ QAH, QSH, and LAF states.  When Zeeman coupling is neglected the Hartree and Fock contributions to the Hamiltonian are mass terms proportional to $\sigma_{\rm z}$, the four flavors decouple, and the mean-field equations are readily solved.  For LAF, QSH, and QAH states the mass terms have the respective forms  $-m\,s_{\rm z}\otimes\sigma_{\rm z}$, $-m\,\tau_{\rm z}\otimes s_{\rm z}\otimes\sigma_{\rm z}$, and $-m\,\tau_{\rm z}\otimes\sigma_{\rm z}$ where $s_{\rm z}$ is a spin Pauli matrix, as summarized in Table~\ref{table:one}.  Using the constant density-of-states per flavor $\nu_0=\gamma_1/(4\pi \hbar^2 v_0^2)$ of the normal state, introducing an ultraviolet cutoff at the inter-layer hopping energy $\gamma_1$, and assuming weak-coupling, the gap equation can be solved to yield
\begin{eqnarray}
m=2\gamma_1 \exp\left({-{2}/{\nu_0 V_{S}}}\right) .
\label{eqn:mass}
\end{eqnarray}

\begin{figure}[t]
\centering{ \scalebox{0.4} {\includegraphics*[-0.0in,2.70in][8.60in,7.85in]{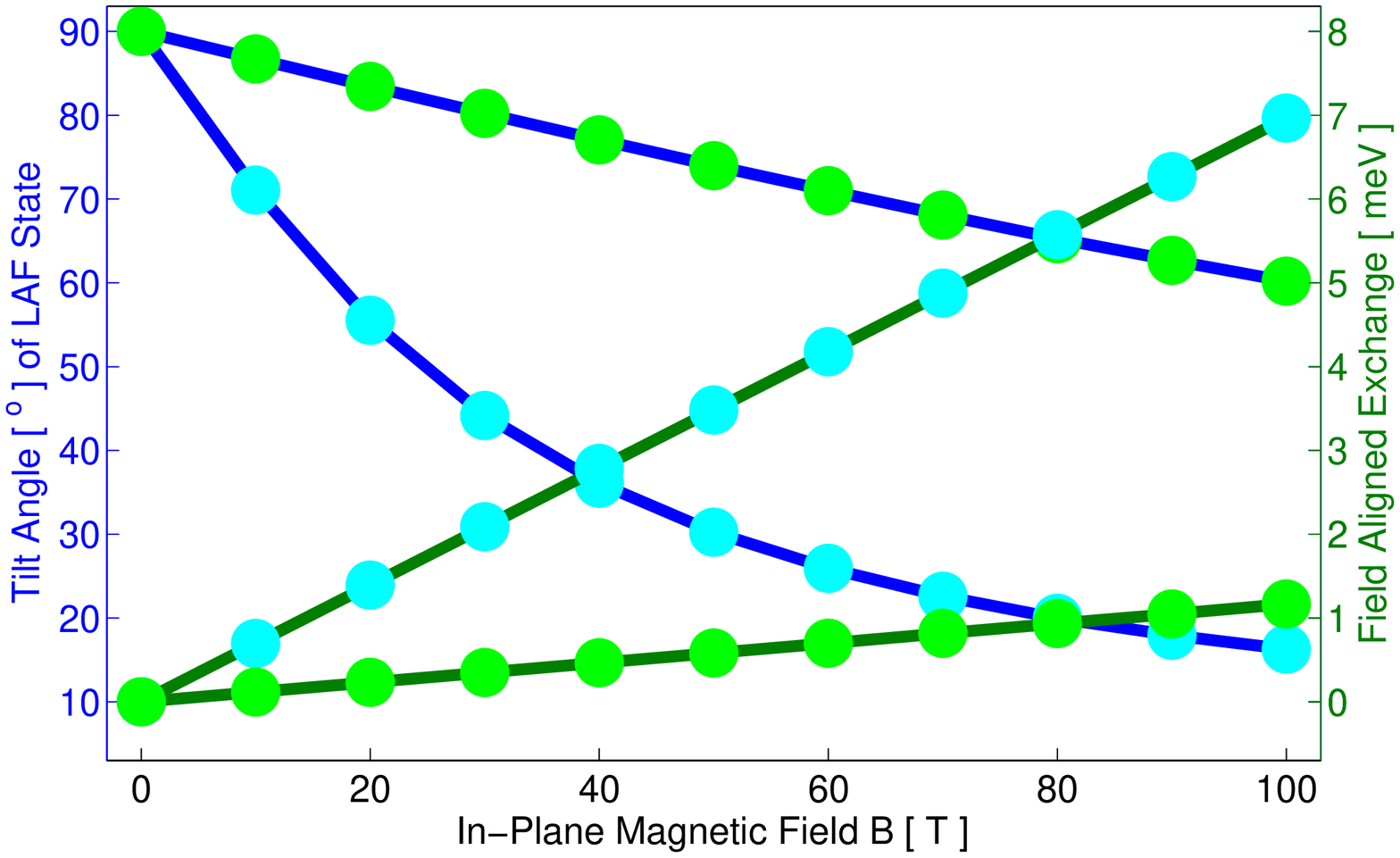}}}
\centering{ \scalebox{0.69} {\qquad\includegraphics*{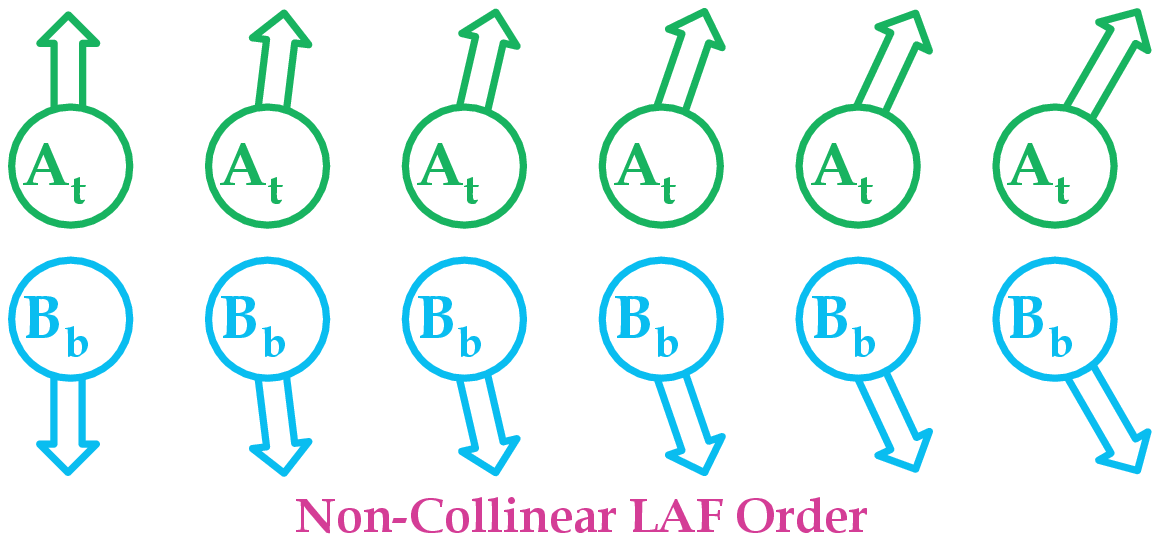}}}
\caption{\label{fig:angle} {(Color online) Upper panel: (Left axis) LAF tilt angle $\theta$ (green) and total effective-field tilt angle including both exchange and external field components (cyan) {\em vs.} in-plane magnetic field.  (Right axis) Field aligned exchange $m\cos\theta$ (green) and total effective-field $M+m\cos\theta$ (cyan) {\em vs.} in-plane field.  Lower panel: sketch of the LAF tilt angles obtained from the upper panel. We assume a $4$ meV spontaneous gap at $B=0$ throughout the paper, corresponding to $\nu_0 V_{S} \sim 0.334$.}}
\end{figure}

{\em Influence of Zeeman Field.}---When Zeeman coupling is included, the QAH state quasiparticles simply spin-split, leaving the ground state unchanged but the charge gap reduced.  For a $4$ meV spontaneous gap at zero-field, corresponding to dimensionless interaction $\nu_0 V_{S} \sim 0.334$ - close to the value expected to be appropriate for screened Coulomb interactions, a field of $\sim 35$ T drives the gap to zero.  The QSH and LAF states, on the other hand, have more interesting non-collinear magnetic-field induced states.  We apply a Zeeman field in the $\hat{x}$ direction and allow spin-densities in the $\hat{x}-\hat{z}$ plane.  In practice this amounts to keeping $\Delta_{\rm \alpha s}^{\rm \beta s'}$ real but allowing spin off-diagonal terms.  In this case we find that for a $4$ meV spontaneous gap, the LAF tilt angle $\theta$ relative to the $\hat{x}$ direction
decreases from $\pi/2$ at zero field to $\pi/3$ at $100$ T.  The mass terms are correspondingly spin-dependent with components in the $\hat{x}$ and $\hat{z}$ directions.  For the LAF
\begin{eqnarray}
h^{\rm HF}_{\rm Z}=h_{0} - m\sin\theta\,s_{\rm z}\!\otimes\sigma_{\rm z}- \left[M+m\cos\theta\right] s_{\rm x}\!\otimes\sigma_{0}\,,
\end{eqnarray}
where $2M=g\mu_{\rm B}B$ denotes the Zeeman splitting and $m$ and $\theta$ are determined by solving
\begin{eqnarray}
\label{eqn:Bmass}
m\sin\theta&=&\frac{V_{\rm S}}{4A}\sum_{\bm k,s=\pm}\frac{m\sin\theta}{E_{\rm s}}\,,\\
\label{eqn:SDWzm}
m\cos\theta&=&\frac{V_{\rm S}}{4A}\sum_{\bm k,s=\pm}\frac{M+m\cos\theta+s \epsilon_{\bm k}}{E_{\rm s}}\,,
\end{eqnarray}
with $E_{\pm}\!\!=\!\!\sqrt{(M+m\cos\theta\pm \epsilon_{\bm k})^2+m^2\sin^2\theta}$.
The four quasiparticle energies are $\pm E_{\pm}$, so the gap is $2 E_{-}$ evaluated at
$\epsilon_{\bm k}=M+m\cos\theta$, {\em i.e.}, $2m\sin\theta$.
\begin{figure}[t]
\centering{ \scalebox{0.39} {\includegraphics*{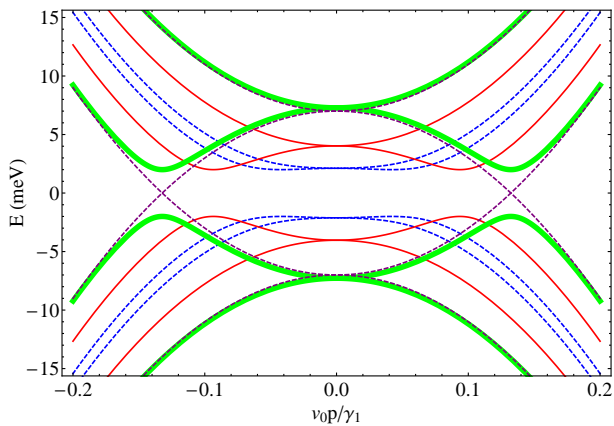}}}
\caption{\label{fig:bands} {(Color online) Canted LAF state quasiparticle bands for a series of in-plane magnetic field strengths:$10$ T (blue), $50$ T (red) and $100$ T (green).  The dashed purple curve for $100$ T shows the quasiparticle bands when the unaligned exchange field $m\sin\theta$ is neglected.}}
\end{figure}

For weak fields the quasiparticle spins are nearly perpendicular to the Zeeman field.  As the field strength is increased the quasiparticle state spin-polarizations, which are $s$ and $\bm k$-dependent, all rotate toward the $\hat{x}$ direction and the exchange field follows suit.  Assuming that $\gamma_1 \gg m, M$ we find that the perpendicular LAF mass component $m\sin\theta$ is still given by the right hand side of Eq.~(\ref{eqn:mass}),
and that
\begin{equation}
m\cos\theta=\frac{\nu_0 V_{\rm S}}{2-\nu_0 V_{\rm S}}M\, ,
\end{equation}
implying that the LAF tilt angle is
\begin{eqnarray}
\theta=\arctan\left[\frac{4\gamma_1\cdot(2-\nu_0 V_{\rm S})}{g\mu_{\rm B}B\cdot\nu_0 V_{\rm S}}\cdot e^{\rm -2/(\nu_0 V_{S})}\right]\,.
\end{eqnarray}
This solution was confirmed numerically and is summarized in Fig~\ref{fig:angle}.

The gap is nearly independent of $M$, in clear contrast to the QAH case.  As $M$ increases the $k=0$ quasiparticle band extrema of the LAF move to larger $k\propto\sqrt{M+m\cos\theta}$ as illustrated in Fig.\ref{fig:bands}. For $M\gg m\sin\theta$ the non-collinear LAF state can be viewed as an exciton condensate formed by pairing electrons in the bilayer majority spin band with holes in the minority spin band. In this limit the LAF state is therefore similar to the Zeeman-coupling induced exciton condensate considered previously in the single-layer graphene case by Aleiner {\em et al.}\cite{Aleiner}.

\begin{figure}[b]
\centering{ \scalebox{0.43} {\includegraphics*[0.15in,2.65in][8.60in,8.05in]{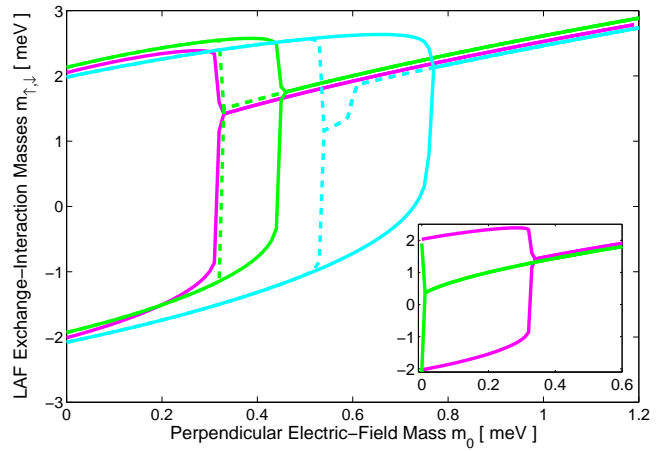}}}
\caption{\label{fig:cdw} {(Color online) LAF exchange-interaction masses $m_{\uparrow,\downarrow}$
at parallel magnetic field $B=$ $0$ T (magenta), $20$ T (green) and $40$ T (cyan) {\em vs.} perpendicular
electric-field mass $m_0$.  The solid and dashed curves were obtained by following the state evolution
{\em vs.} electric field at fixed Zeeman field and {\em vs.} Zeeman field at fixed electric field.
The inner panel indicates the LAF (magenta) and fully layer polarized (green) state stability ranges
{\em vs.} electric-field mass at zero magnetic field.}}
\end{figure}

{\em Influence of Electric Field.}---Because they all have $\sigma_{z}$ layer pseudospin order, LAF, QAH, and QSH states respond similarly to an electric field perpendicular to the layers, which adds a $m_0\sigma_{\rm z}$ term to the single-particle Hamiltonian.  For the LAF, for example, the LAF masses $m_{\uparrow,\downarrow}$ for $m_0=0$ differ only by a sign.  When a perpendicular electric field is applied, masses are enhanced for one spin and suppressed for the other.  In our mean-field calculations first order phase transitions occur between states with distinct broken symmetries
as illustrated in Fig~\ref{fig:cdw}, leading eventually to a state in which the sense of layer polarization is the same for all spin-valleys\cite{Tutuc}.  Experimental behavior in an external electric field will likely be sensitive to the pinning energies of domain walls that separate different spontaneous quantum Hall states.

When an in-plane magnetic field and a perpendicular electric field are both present, the field aligned LAF order parameter $m\cos\theta$ is little changed compared to the $E=0$ case.  The electric field dependence of $m_{\uparrow,\downarrow}$ is mainly determined by a competition between $m_0\sigma_{\rm z}$ and $m\sin\theta\,s_{\rm z}\!\otimes\sigma_{\rm z}$.  The noncollinear LAF phase is, however, strengthened by its field-aligned order-parameter component and is more robust against a perpendicular electric field when the Zeeman field is large, as illustrated in Fig~\ref{fig:cdw}.  Fig~\ref{fig:cdw} also shows that the LAF state stability can be dependent on the order in which the two fields are applied.
We note that a small electric field between the layers can stabilize a state in which one flavor is polarized in a sense opposite to the other three and charge, valley, and spin Hall conductivities are all non-zero\cite{SQH_Zhang}.  This state is not
represented in Fig.~\ref{fig:cdw} where we have assumed that the two valleys have the same layer polarization.

{\em Discussion.}---Low-energy electrons in bilayer graphene have spin, valley, and layer two-component quantum degrees of freedom.  Because it appears in the band Hamiltonian, the layer pseudospin plays a different role in bilayer graphene physics than spin or valley.  Flat conduction and valence bands and Bloch states with $J=2$ layer-pseudospin chirality combine  to make the band state unstable toward a family of insulating broken symmetry states that have independent spontaneous layer polarizations in each spin-valley component.  Three distinct states have no overall layer polarization, a quantum anomalous Hall state, a quantum spin Hall state, and a layer antiferromagnet state\cite{SQH_Zhang}.  In this Letter we have shown that the QAH state can be distinguished from the QSH and LAF states by examining the dependence of the charged quasiparticle gap on the strength of Zeeman coupling to an in-plane magnetic field.  In the QAH case, the ground state is unchanged but the quasiparticle gap is reduced - vanishing when the Zeeman coupling strength is equal to the ground state gap via a mechanism reminiscent of the Clogston limit in superconductors.
The QSH and LAF states respond to Zeeman fields in a more interesting way, by establishing non-collinear spin states within each valley and evolving toward an unusual kind of exciton condensate in the strong Zeeman coupling limit.  The gap of QSH and LAF states is independent of Zeeman coupling strength drawing a sharp distinction with the QAH case.  When combined with probes that are sensitive to edge state transport, which is topologically protected\cite{TI} in QAH and QSH cases but not in the LAF case, this property should enable any of the three states to be uniquely identified.

It appears clear that bilayer graphene is exhibiting new many-body physics.  This Letter points out that experimental studies of the Zeeman energy dependence of the gap could help to distinguish between different possibilities in bilayers, and also in larger $N$ chiral few-layer graphene.
As mentioned previously\cite{ft2} some theoretical authors have concluded\cite{RG_Vafek,RG_Falko} that the ground state of a neutral bilayer should be a {\em nematic} $XY$-plane layer-pseudospin ferromagnet which breaks in-plane rotational symmetry, rather than a $\hat{z}$-direction Ising pseudospin
ferromagnet.  (The z-component of the layer pseudospin density is the difference in density between the top and bottom layers while an x- or y-component indicates interlayer coherence.)  The nematic states are most strongly distinguished from the  \,$\mathcal{I}$\,-symmetry breaking  spontaneous quantum Hall states\cite{SQH_Zhang,MF_Min,RG_Zhang,MF_Levitov,SQH_Levitov,MF_Jung} by the absence of a charged quasiparticle gap in the former case.  In the nematic state interactions generate mean fields that are off-diagonal in layer index and reduce the symmetry of the bands, splitting the $2 \pi$ $K (K')$ Dirac points into two $\pi$-Dirac points that are displaced from $K (K')$ in an arbitrary direction.  The mean-field-theory property that lower energy
states are obtained with Ising compared to $XY$ pseudospin order is related to the larger susceptibility associated with this
pseudospin component.  (The band eigenstates are perpendicular to the $\hat{z}$-direction for all ${\bm k}$, so all band states are easily rotated toward $\hat{z}$ pseudospin polarization.)
Other potential explanations for the anomalies observed to date can be sought in trigonal warping effects, which are relevant  below $\sim 1$ meV in bilayers and have been ignored for simplicity in the present discussion, and
in structural changes unintentionally induced by current annealing of suspended samples.
There is however not yet a coherent explanation of
how either of these might result in a gap at Dirac point.
The observed gaps appears to be of many-body origin, in any event, since
they appear only at temperatures that are much lower than
observed gaps\cite{Yacoby2,AB_Schonenberger,AB_Lau,AB_Lau1,ABC_Lau}.

{\em Acknowledgement.}---This work has been supported by Welch Foundation under Grant No. TBF1473, NRI-SWAN, DOE Division of Materials Sciences and Engineering Grant No. DEFG03-02ER45958, NSF under Grants No. DMR-0606489 and No. DMR-0955778, and ARO W911NF-09-1-0527. We acknowledge helpful discussions with C. Lau, B. Halperin, A. Yacoby, K. Novoselov, W. Bao, J. Velasco, D. Tilahun and J. Jung.


\begin{thebibliography}{100}
\bibitem{ft1} {We assume chiral (ABC) stacking throughout this paper.}
\bibitem{ft2}
Some authors have concluded that the ground state is an XY layer-pseudospin ferromagnet.  The XY pseudospin ferromagnet is anisotropic and is therefore usually referred to as a nematic state.  This state has no charge gap and no momentum space Berry curvature.
\bibitem{ft3}
 We adopt this language. However, because valley is not rigorously a good quantum number,
 the experimental properties of the various types of possible spontaneous Hall states have to be
 examined carefully.
\end{thebibliography}
\end{document}